# Recent progress on thermal transport in one-dimensional long-range interacting Fermi-Pasta-Ulam-Tsingou lattice systems*

XIONG Daxing[1, *], LI Nianbei[2, *], CHEN Jie[3, *]

1. MinJiang Collaborative Center for Theoretical Physics, College of Physics and Electronic Information Engineering, Minjiang University, Fuzhou 350108, China
2. Institute of Systems Science, Department of Physics, College of Information Science and Engineering, Huaqiao University, Xiamen 361021, China
3. MOE Key Laboratory of Advanced Micro-Structured Materials, China-EU Joint Laboratory for Nanophononics, Center for Phononics and Thermal Energy Science, School of Physics Science and Engineering, Tongji University, Shanghai 200092, China

**Abstract** Long-range (LR) interactions are no longer just a theoretical idea: they can be engineered in several low-dimensional platforms and offer a new way to control heat transport. At the same time, they push beyond the standard picture developed mainly for short-range, momentum-conserving lattices. In this review, we consider one-dimensional Fermi-Pasta-Ulam-Tsingou (FPUT)-type lattices where LR effects enter mainly through a quartic anharmonic coupling that decays as a power law, characterized by an exponent $\sigma$. Our goal is to explain, in a unified way, how LR anharmonicity changes microscopic energy exchange, collective dynamics, and the macroscopic scaling laws of heat conduction-while also clarifying what is well established and what is still debated.

We first introduce the main models and the quantities used to discribe transport in LR-FPUT systems. These include the nonequilibrium steady-state thermal conductivity $\kappa(N)$, the equilibrium Green-Kubo (GK) approach based on the heat-current autocorrelation $C_{JJ}(t)$, and spatiotemporal correlations of heat/energy fluctuations $\rho_Q(m,t)$, together with the scaling of the mean-squared displacement (MSD). A key technical point is that, in LR lattices, the microscopic heat current contains explicit nonlocal contributions. As a result, different (formally related) implementations of the current may lead to noticeable differences at finite system sizes and finite times. To extract the intrinsic transport behavior, we therefore stress practical checks: i) the stability of scaling exponents against boundary and driving protocols, ii) validation among all the three methods, i.e., $\kappa(N)$, $C_{JJ}(t)$, and $\rho_Q(m,t)$, and iii) careful control of finite-size and finite-time crossovers.
Based on these methods, we review representative $\sigma$-dependent transport regimes. i) The special point$\sigma = 2$ is central and remains controversial. Early work reported ballistic-like behavior under certain boundary conditions. More recent studies using more controlled numerical method (e.g., periodic driving together with correlation scaling) tend to support strongly enhanced superdiffusion, with an effective divergence exponent around $\alpha \simeq 0.7$, rather than ballistic transport. The most consistent physical picture links this enhancement to weak nonintegrability and the presence of long-lived coherent carriers, such as mobile breathers or related nonlinear excitations, while recognizing that finite-size results are unusually sensitive to boundary implementation and to how the heat current is defined. We summarize the current consensus and the main remaining disagreements, and we argue that ballistic vs strong superdiffusion should be decided by consistency across multiple observables, not by temperature profiles alone. ii) In the weak-LR regime ($1 \leqslant \sigma \leqslant 3$), a striking result is a near-diffusive window around $\sigma \simeq 1.25$. This is conceptually important because it suggests a possible route toward quasi-normal transport within Hamiltonian, momentum-conserving one-dimensional lattices, which usually show anomalous conduction in the short-range case. We discuss how this challenges the common intuition that momentum conservation implies anomalous transport, but we also emphasize that the current evidence should be viewed as a strong candidate that still needs careful confirmation against finite-size crossovers and method dependence. iii) In the strong-LR regime ($0 \leqslant \sigma \leqslant 1$), simulations show clear antipersistent (negative) structures in $C_{JJ}(t)$ and a trend toward slower transport, sometimes described as subdiffusive-like transport. iv) We also discuss an inverse LR coupling design LR-FPUT model that can invoke ballistic transport, suggesting that LR nonlocality combined with structural modulation may open new transport channels and may be useful for thermal rectification.
Finally, we outline key open directions: establishing reliable finite-size scaling to distinguish genuine $\sigma$-driven dynamical transitions from broad crossovers; mapping transport regimes in a $(\sigma, T, \text{nonlinearity})$ phase diagram; and identifying the microscopic carriers (phonons, stationary/mobile breathers) and their scattering dynamics that underpin the observed superdiffusive, near-diffusive, and antipersistent slow-transport behaviors. These needs also connect to experiments, since LR couplings can be realized in platforms such as Coulomb crystals and frustrated magnets, and LR interaction engineering has already enabled efficient thermal rectification-suggesting practical routes to tune heat conduction and build thermal diodes/switches based on nonlocality and nonlinearity.

# 1 Introduction

Heat transport in one-dimensional (1D) systems serves as a classic example for understanding macroscopic irreversible laws from microscopic Hamiltonian dynamics. For nearest-neighbor nonlinear lattice systems, theoretical and numerical studies have demonstrated that if momentum is conserved, the thermal conductivity $\kappa$ typically diverges as a power law $\kappa \sim N^{\alpha}(0<\alpha<1)$ with system size, exhibiting anomalous heat conduction. Conversely, if momentum is not conserved (such as in systems with external fields or on-site potentials), the system reverts to normal heat conduction satisfying Fourier's law, where $\kappa$ remains independent of system size in the thermodynamic limit. This framework has been systematically summarized in relevant review articles[1-7] and has become fundamental knowledge in the field of low-dimensional heat transport.

Research on this fundamental problem in statistical physics has a long history. As early as 1967, Rieder et al.[8] investigated heat transport in a 1D integrable harmonic oscillator lattice heated at both ends. Their theoretical analysis revealed that the heat flux in the system remains constant. Combined with Fourier's law, this implies that the thermal conductivity $\kappa$ increases linearly with system size $N$, following $\kappa \sim N$. This corresponds to ballistic transport in transport theory, manifesting as the physics of non-interacting phonon ballistic transport. Another typical integrable system is the Toda lattice[9], which exhibits soliton dynamics that preserve both energy and momentum. Subsequently, in 1984, Casati et al.[10], pioneers in classical and quantum chaos, designed and studied a 1D lattice chain model with chaotic characteristics. They observed normal heat transport in the 1D case, suggesting that non-integrable chaotic motion is a necessary and sufficient condition for the validity of Fourier's law. However, thirteen years later in 1997, Lepri et al.[11] challenge this view. When studying the 1D nonlinear Fermi-Pasta-Ulam-Tsingou-$\beta$ (FPUT-$\beta$) model, which also exhibits typical chaotic characteristics, they discovered superdiffusive transport following $\kappa \sim N^{\alpha}$. This corresponds to phonon-phonon interactions arising from the introduction of nonlinearity, where the anomalously slow decay of long-wavelength phonons leads to superdiffusive transport intermediate between normal and ballistic regimes. In 1998, Hu et al.[12] further designed and studied a class of Frenkel-Kontorova models and $\phi^4$ models with on-site potentials, explicitly identifying momentum conservation as a key factor determining whether low-dimensional systems obey normal transport.

In fact, the relationship between transport and system conserved quantities resembles predictions from classical fluctuating hydrodynamics developed by Ernst et al.[13] in the 1960s and 1970s. Researchers recognized this connection and, in 2002, explained the anomalous transport law $\kappa \sim N^{\alpha}$ from a hydrodynamic perspective using hydrodynamic renormalization group methods[14]. However, it was not until 2012 and 2014 that van Beijeren[15] and Spohn[16] respectively established corresponding nonlinear fluctuating hydrodynamics theories to predict heat transport laws in 1D lattice systems. The core idea focuses on the three conserved quantities generally present in systems exhibiting superdiffusive transport: momentum, particle density (or elongation), and energy. The correlation matrix of hydrodynamic modes is derived via canonical transformation from the spatiotemporal correlation functions of these conserved quantities, providing detailed transport information. Breaking momentum (or elongation) conservation leaves energy as the sole conserved quantity, leading to normal transport, as exemplified by rotor models[17,18]. In recent years, Doyon et al.[19] have been further developing the theory of generalized hydrodynamics, aiming to understand ballistic transport in integrable systems possessing infinitely many conserved quantities.

The tremendous success of hydrodynamic theory has often led researchers to overlook kinetic effects in transport theory (refer to the trilogy proposed by Bogoliubov in 1946[6]). Recently, inspired by new advances in the study of classical thermalization problems in transport theory[20-24], Professor Zhao Hong's research group[5,25] and the Italian research groups of Lepri and Livi[26] have successively proposed that a complete picture of energy transport in 1D lattices must consider both kinetic and hydrodynamic effects. Combined with recent progress in heat transport in 1D spin chains[27-29], these results have further deepened the understanding of energy transport theory in long-range(LR) lattices.

On the other hand, systems with long-range interactions have long been a research hotspot in statistical physics and non-equilibrium dynamics. These systems span multiple scales, from gravity-dominated celestial systems to nanoscale materials and mesoscopic structures. A general characteristic of LR systems studied by physicists is that the interaction potential decays as a power law with the interparticle distance *r*:

$$V(r) \propto 1/r^{\sigma}, \qquad (1)$$

Here, $\sigma$ denotes the LR decay exponent. This leads to thermodynamic and dynamic properties that differ significantly from those of nearest-neighbor and short-range systems. Examples include non-ergodicity, weak chaos, ensemble inequivalence, long-lived non-Gaussian states, 1D phase transitions, non-concave entropy, and even negative specific heat[30-35]. These phenomena challenge the classical Boltzmann-Gibbs statistical framework and motivate the development of more general LR non-equilibrium statistical theories.

However, although one might intuitively expect the interaction range to significantly affect heat transport, mainstream theoretical understanding of heat transport has long relied on lattice models with nearest-neighbor short-range coupling. Only in recent years has heat transport in LR interacting lattice systems been systematically studied. The challenges in studying heat transport in such systems are twofold. From a dynamical perspective, when interactions extend beyond nearest neighbors to non-local couplings that decay as a power law with distance, the phonon dispersion of the system changes. This introduces nonlinear scattering channels and long-time memory structures. Consequently, traditional understanding of heat transport in nearest-neighbor short-range systems, which is based on local conservation laws, may require revision. From a methodological perspective, LR coupling introduces difficulties in defining boundary conditions and heat flux in non-equilibrium steady-state simulations. This affects the determination of intrinsic transport properties. To address the latter issue, researchers have proposed combining periodic boundary conditions with reverse non-equilibrium molecular(RNEMD) dynamics. This approach overcomes these difficulties and reveals heat transport characteristics in one-dimensional LR interacting FPUT lattices that are closer to reality than those observed under traditional fixed boundary conditions[36].

Indeed, various one-dimensional LR interacting lattice models analogous to short-range systems have been studied in this emerging field. These include LR rotor chains, LR FPUT chains, LR $\phi^4$ chains, and LR harmonic chains[36-50]. It is important to emphasize that many of these models preserve momentum conservation. Therefore, they provide a more general platform for testing the traditional view that momentum conservation necessarily leads to anomalous heat conduction.

Take the LR rotor chain as an example. Previous studies indicate that the system undergoes a transition from an insulating phase to a conducting phase as the LR decay exponent $\sigma$ varies[37]. More interestingly, in the strong LR regime ($0 < \sigma < 1$), the system exhibits a nearly flat temperature profile. This behavior superficially resembles that of integrable systems. However, subsequent analysis suggests that this likely arises from a parallel energy transport mechanism rather than true integrability[40].

Even more notable is the LR-FPUT system, where a key point of repeated discussion occurs at $\sigma = 2$. In 2017, Bagchi[38] demonstrated that the thermal conductivity grows nearly linearly with system size and the temperature profile is nearly flat. These features indicate ballistic-like transport and even sparked conjectures about a hidden integrable limit. However, more detailed analyses of boundaries and thermal baths yielded different conclusions. The system is clearly non-integrable and exhibits superdiffusive heat transport[36]. This conclusion is supported by further evidence, such as slower decay of energy current correlations, weaker signatures of non-integrability, and the emergence of specific carriers, namely moving breathers, which enhance heat transport at $\sigma = 2$. The controversy surrounding the LR-FPUT system at $\sigma = 2$ highlights that heat transport in LR interacting lattice systems may differ substantially from that in traditional nearest-neighbor lattice systems.

In this context, this paper systematically reviews several distinctive new findings regarding 1D LR interacting FPUT lattice systems:

1) Weak non-integrability and significantly enhanced anomalous heat conduction occur at a specific LR decay exponent ($\sigma = 2$). This phenomenon is related to the dynamics of moving breathers in the system;

2) Energy subdiffusion and negative correlations in the energy current autocorrelation function are observed in the Hamiltonian mean-field system;

3) In the strong LR regime ($0 < \sigma \leqslant 1$), the energy current autocorrelation exhibits a complex negative correlation structure. This structure varies non-monotonically with $\sigma$ and undergoes a significant transition at $\sigma_c = 0.5$;

4) A window of normal heat transport approaching normal diffusion appears in the weak LR regime ($1 < \sigma \leqslant 3$), specifically near $\sigma_c = 1.25$;

5) All characteristics of ballistic heat transport are observed in a non-integrable classical long-range FPUT-like system with staggered LR inverse coupling. This indicates that non-integrable LR lattice systems can exhibit ballistic heat transport, a feature typically associated only with integrable systems in short-range nearest-neighbor contexts.

# 2 Models and Numerical Methods

This paper primarily focuses on 1D LR interacting FPUT lattice systems. The Hamiltonian is expressed as

$$H = \sum_{i=1}^{N} [\frac{p_i^2}{2} + \frac{1}{2}(x_{i+1} - x_i)^2 + \frac{1}{4}\sum_{r=1}^{N/2-1} \frac{(x_{i+r}-x_i)^4}{r^\sigma}]. \quad (2)$$

This study employs the Born-von Karman periodic boundary conditions from solid-state physics. In the Hamiltonian, $x_i$ and $p_i$ denote the displacement from equilibrium and the momentum, respectively, of the particle labeled *i*. *N* represents the number of particles. By convention, the atomic mass and lattice constant are dimensionless and set to unify, so *N* also indicates the system size. This system exhibits three characteristics: 1) It lacks external fields and potentials, ensuring conservation of total momentum. Based on conventional understanding of short-range nearest-neighbor systems, anomalous heat conduction is typically expected. 2) The harmonic terms remain nearest-neighbor, indicating that the linear spectrum remains short-range and low-frequency acoustic modes are not directly long-ranged. 3) LR interactions manifest in the quartic anharmonic terms, which effectively modify the scattering kernel and energy scattering channels.
It is crucial to emphasize that quartic anharmonic terms correspond to four-wave interactions in the normal mode (phonon) representation. In the context of the crystal phonon Boltzmann transport equation, they correspond to the fourth-order force constants governing four-phonon scattering. Therefore, the discussion herein regarding "LR quartic anharmonic coupling altering scattering channels" naturally parallels recent first-principles studies on "four-phonon scattering dominance" or "strain-regulated transitions in scattering mechanisms"[51,52]. Within this theoretical framework, Reference [53] demonstrates that for the $\sigma = 2$ system under this Hamiltonian, the quartic anharmonic terms exhibit a specific potential function symmetry in Fourier space due to four-phonon scattering processes. This symmetry allows the $\sigma = 2$ system to excite moving breathers without tails, leading to its unique characteristics. This feature may significantly influence the transport behavior of the system.
Furthermore, existing studies implement LR interactions in FPUT –systems primarily through two approaches: making only the quartic anharmonic terms long-range, or making both quadratic and quartic terms LR. Many studies also introduce the Kac scaling factor[33] to ensure the extensivity of the system Hamiltonian. This paper focuses on systems where only the quartic anharmonic terms are long-range and no Kac scaling factor is applied. We adopt this approach to prevent the long-range modification of low-frequency acoustic mode structures. Additionally, the Kac factor causes phonon group velocity and nonlinear strength to depend on system size, which is undesirable for investigating the dependence of thermal conductivity on system size in heat transport studies. Moreover, introducing the Kac factor significantly prolongs the time scale required for the system to reach equilibrium.
Effective theoretical analytical tools for studying heat transport in long-range interaction systems are currently lacking, making numerical calculations essential. Fortunately, most traditional methods used for numerically studying short-range nearest-neighbor systems remain applicable. Here, we elaborate on the advantages and disadvantages of these methods when applied to LR interaction systems.
1) Non-equilibrium molecular dynamics (NEMD)[1,2]. This method establishes a steady-state heat flux $J$ by connecting the two ends to heat baths at different temperatures over an extended period. The divergence exponent $\alpha$ is then derived using Fourier's law, $\kappa(N) = J/\nabla T$ (where $\nabla T$ is the temperature gradient, examining $\kappa \sim N^{\alpha}$). A challenge with this method is its general requirement for fixed boundary conditions. However, introducing fixed boundary conditions may have a non-negligible impact on the heat transport behavior of LR systems, thereby preventing the observation of intrinsic heat transport properties.

2) Green-Kubo formula method[1,2]. This method calculates the autocorrelation function of the energy flux in the equilibrium state:

$$C_{JJ}(t) = \langle J_{\mathrm{tot}}(t) J_{\mathrm{tot}}(0) \rangle, \qquad (3)$$

The long-time tail phenomenon of this correlation function is closely related to the transport behavior of the system. However, in LR systems, the expression for heat flux often includes explicit LR contributions. This creates significant computational difficulties, preventing the system sizes studied in LR systems from reaching those achievable in short-range nearest-neighbor systems. Specifically, in lattices with LR interactions, the explicit expression for heat flux (or energy flux) includes not only pairwise transmission terms spanning long distances but also depends on how potential energy is partitioned among lattice sites. Consequently, different studies may adopt different heat flux expressions in their numerical implementations under various boundary conditions, although these expressions are formally equivalent or approximately equivalent. In finite systems and within finite time windows, such differences may alter the long-time behavior of $C_{JJ}(t)$. They may even lead to deviations in the quantitative interpretation of $\alpha$ extrapolated from $\kappa \sim N^{\alpha}$. To reduce such deviations, one can simultaneously cross-validate the long-time behavior of the Green-Kubo $C_{JJ}(t)$, the divergent behavior of the non-equilibrium $\kappa(N)$, and the scaling properties of the spatiotemporal correlation function of thermal energy density fluctuations described in method 3) below. This approach serves as a robust consistency check for investigating the transport properties of the system.

3) Spatiotemporal correlation function scaling method for thermal energy density fluctuations[54]. This method calculates the spatiotemporal correlation function of heat (or energy) density fluctuations:

$$\rho_Q(m,t) = \frac{\langle \Delta Q_{l+m}(t) \Delta Q_l(0) \rangle}{\langle \Delta Q_l(0) \Delta Q_l(0) \rangle}, \qquad (4)$$

In the equation, $\langle \cdot \rangle$ denotes the ensemble average; $Q_l(t) = E_l(t) - [(\langle E \rangle + \langle F \rangle) g_l(t)]/\langle g \rangle$ represents the thermal energy density of the system within a specific coarse-grained spatial bin $l$. Its value depends on the energy $E_l(t)$, particle number $g_l(t)$, and force $F_l(t)$ within the same bin at time $t$. After obtaining $\rho_Q(m,t)$ through numerical calculations, analyzing its scaling behavior (such as Lévy-type or non-Gaussian long tails) allows for the differentiation of various thermal diffusion behaviors and the identification of corresponding universality classes. This approach is considered a refined spatiotemporal scaling analysis for investigating thermal transport.

Furthermore, the primary challenge in numerically simulating LR systems lies in the computational complexity of calculating LR forces during numerical integration. Fortunately, fast Fourier transform algorithms(FFT) for 1D lattices with LR interactions have been developed, effectively accelerating computational processes in integration. To address difficulties in defining boundary conditions and heat flux when using heat baths at both ends in traditional non-equilibrium molecular dynamics methods for LR systems, some studies have proposed the " RNEMD " method[36]. The core concept of this method differs from traditional approaches that directly induce temperature gradients. Instead, it imposes a heat flux by frequently exchanging the kinetic energy (or momentum) of particles. A temperature gradient is established once the non-equilibrium steady state is reached. This "reversal" makes the method an ideal choice for studying thermal transport in systems with LR interactions. We also note that VASP and LAMMPS codes implementing this method are widely used. The advantage of this approach is that it is a molecular dynamics method naturally developed under periodic boundary conditions for thermal conduction studies, effectively avoiding finite-size effects associated with fixed boundary conditions. Additionally, the definition and acquisition of heat flux occur naturally in this method, making it well-suited for investigating thermal transport in LR interaction systems.

Finally, this paper examines common numerical methods for studying thermal transport in LR FPUT systems. A brief comparison of these methods is summarized in Table 1. It should be noted that most numerical results reviewed in this paper were obtained under conditions of fixed average temperature (mostly $T = 0.5$) and given nonlinear strength parameters. Given that the transport exponent $\alpha$ may vary with temperature (energy density) in short-range FPUT chains, the influence of temperature $T$ and nonlinear strength on transport phenomena in LR FPUT systems remains a direction for further research and is not discussed in this review.

**Table 1.** Comparison of commonly used numerical methods for investigating thermal transport in LR-FPUT systems (accuracy, efficiency, and system-size limitations).

| Method | Critical output | Advantage (long-range system) | Primary Risk/Source of Deviation | Computational Cost and Size Constraints |
|---|---|---|---|---|
| Two-terminal heat-bath NEMD (fixed boundary) | $J,\ \nabla T,\ \kappa(N)$ | Intuitive; $\kappa(N)$ scaling obtained directly | Boundary layer thickness; heat bath/heat flux definition sensitive; small $N$ easily leading to "ballistic-like" | $O(N^2)$; FFT can reach $O(N\log N)$; reaches steady state slowly |

| Method | Critical output | Advantage (long-range system) | Primary Risk/Source of Deviation | Computational Cost and Size Constraints |
|---|---|---|---|---|
| RNEMD (periodic boundary) | Applied heat flow (exchanged kinetic energy) $\nabla T, \kappa(N)$ | Cleaner body regions; reduced endpoint artifacts; more robust | Crossover introduced by exchange parameters; statistical noise requires long averaging | Limited by long-range forces; often requires long time averaging and large $N$ |
| Green-Kubo (equilibrium state) | $C_{JJ}(t),\ \kappa \propto \int C_{JJ}$ | Heat-bath-free boundary; Tail/Sign Change/Anti-persistence in Decidable Long | The heat flux expression is complex and implementation difference, and is sensitive to the truncation of the long time tail | Long trajectories and multiple samples are required, and the convergence cost is high for the slow-tail |
| Spatiotemporal correlation ( $\rho_Q$ , MSD) | Morphology (Gaussian/heavy-tailed), MSD index $\beta$ | Intuitively identifies diffusion regimes; less sensitive to the form of heat flux | Rewind/touch edge to be avoided; cross drift in exponential fit | The cost is limited by the long-range force; the shape of the is easy to be stable, and the exact exponent needs a large $N$ |

# 3 Overview of Representative Results

## 3.1 Thermal Transport Puzzle at $\sigma = 2$: Ballistic vs. Anomalous

In studies of thermal transport in 1D LR FPUT systems, the case of $\sigma = 2$ initially attracted significant interest. Using NEMD, Bagchi[38] first reported in 2017 that LR FPUT systems at $\sigma = 2$ exhibit thermal transport behavior approaching ballistic transport, a phenomenon traditionally observed only in integrable systems. The most intuitive evidence supporting this finding is that the thermal conductivity $\kappa$ increases linearly with system size, while the temperature profile becomes flat except at the heat source boundaries. Subsequently, Iubini et al.[40] considered the specific characteristics of LR systems. They employed NEMD with a slightly different heat bath configuration than Bagchi and carefully distinguished between intrinsic energy flux and heat-bath-driven energy flux. Their work largely confirmed the quasi-ballistic thermal transport behavior observed by Bagchi. These observations sparked speculation about the existence of a "hidden integrable limit" in LR FPUT systems at $\sigma = 2$.

Based on the premise that ballistic transport cannot occur in non-integrable systems, Wang et al.[36] pointed out that the nonequilibrium steady-state drive in LR systems is strongly influenced by two factors when using traditional nonequilibrium molecular dynamics: 1) boundary conditions and heat bath coupling methods; and 2) the definition of heat flux. To extract transport properties "closer to reality," Wang et al. used RNEMD to establish the steady-state drive, thereby avoiding strong boundary artifacts potentially caused by conventional heat bath schemes under LR coupling. Under this setup, they observed a power-law divergence of thermal conductivity with size, $\kappa \sim N^{\alpha}$, at $\sigma = 2$. They obtained a significantly large divergence exponent $\alpha \approx 0.71$ (see Figure 1), indicating that the system is in a state of strong superdiffusive transport. This state is markedly stronger than the anomalous transport commonly seen in short-range nearest-neighbor chains but does not reach the strict ballistic limit ($\alpha = 1$). The distinctly non-flat temperature gradient distribution shown in Figure 2 also supports this finding.

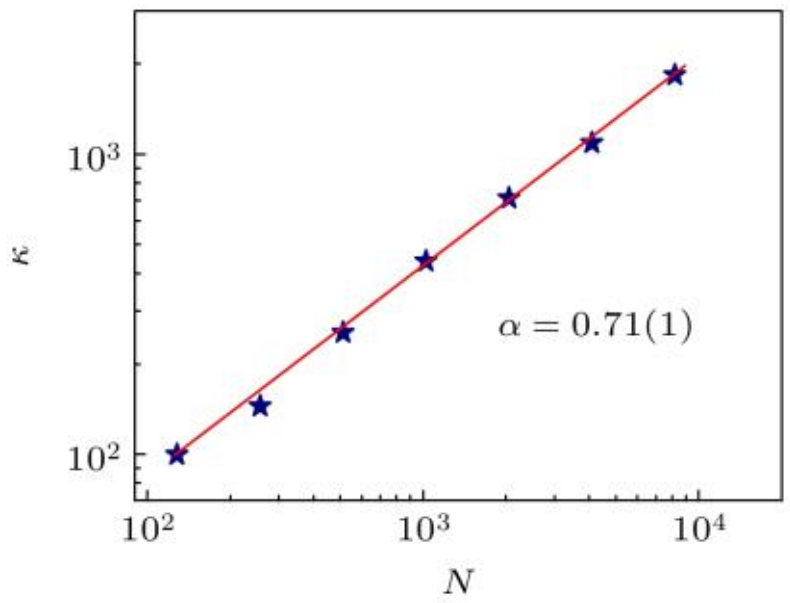

**Fig. 1 The dependence of heat conductivity $\kappa$ on system size $N$ in the case of $\sigma = 2$ (the average temperature here is $T = 0.5$, the same below)**

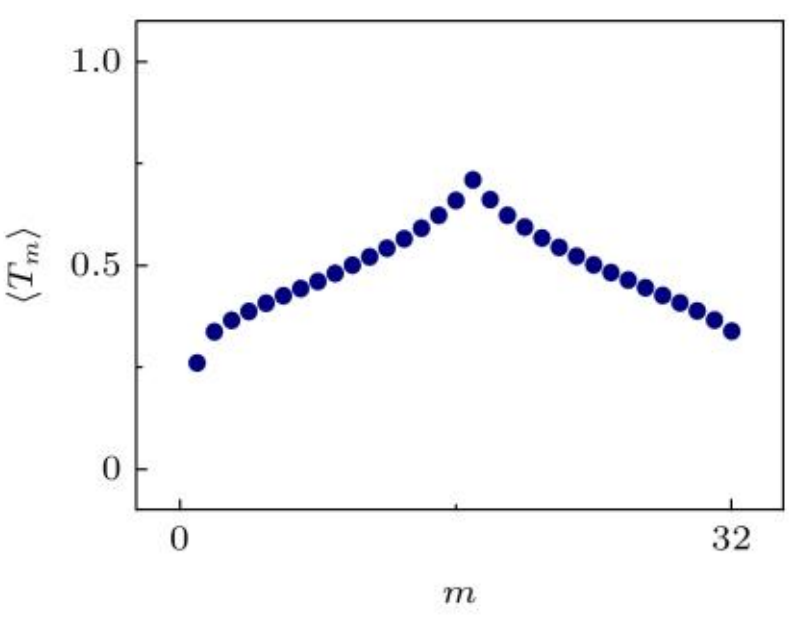

**Fig. 2 The temperature profile of the $\sigma = 2$ system.**

The significance of this result lies in the fact that LR interactions do not merely fine-tune the exponent of short-range anomalous heat conduction. Instead, they may drive the system into a new scaling regime. Consequently, under Hamiltonian dynamics that also conserve momentum, a stronger divergence exponent for heat conduction can be achieved.

Regarding the underlying microscopic mechanism, anomalous heat conduction generally implies that microscopic heat carriers follow anomalous energy or heat diffusion. This is reflected in the spatiotemporal correlation function of heat energy density fluctuations, $\rho_Q(m,t)$, which conforms to the spatiotemporal scaling characteristics of anomalous transport, namely

$$t^{1/\gamma}\rho(m,t) \approx \rho(t^{-1/\gamma}m,t). \qquad (5)$$

Reference [36] further investigated $\rho_Q(m,t)$ and its scaling properties (see Figure 3). The study identified a novel distribution function featuring a persistent intermediate plateau. It also reported a spatiotemporal scaling exponent of $\gamma \approx 1.29$ $(1/\gamma = 0.78)$. Interestingly, applying the formula $\alpha = 2-\gamma$ from Lévy walk theory[55], which links $\alpha$ and $\gamma$, yields $\alpha = 0.71$. This result aligns with anomalous heat conduction, thereby supporting fundamental insights into anomalous diffusion.

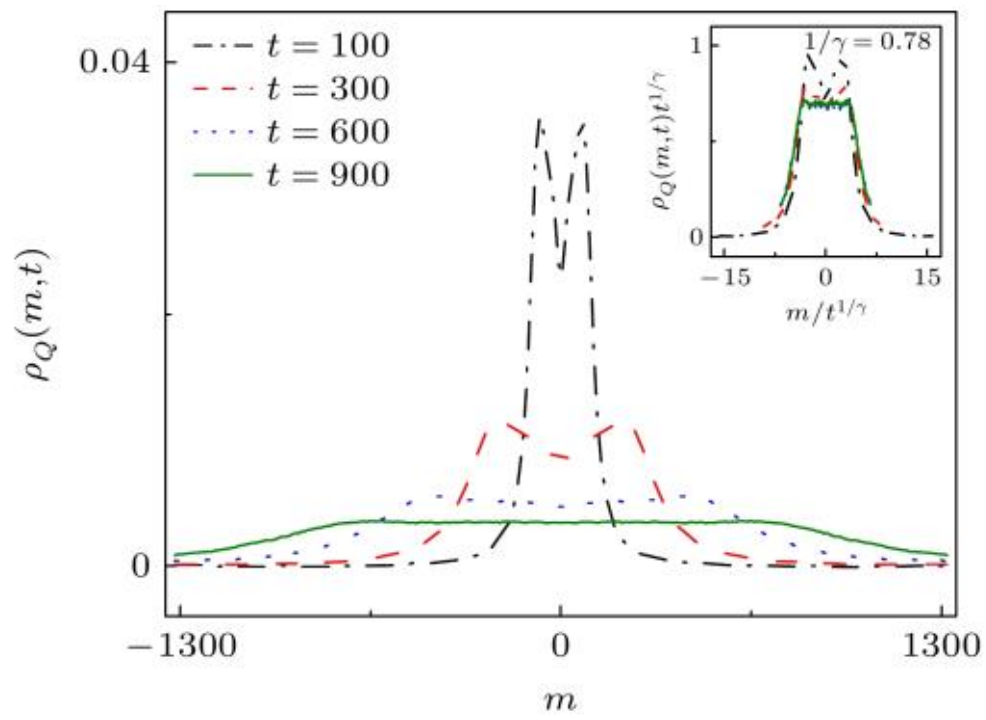

**Fig. 3 The $\rho_Q(m,t)$ of the $\sigma = 2$ system at different times, the inset shows its scaling behavior.**

Reference [36] attributes this novel anomalous heat transport behavior to the weak non-integrability of the system and the unique dynamics of microscopic moving breathers. Regarding weak non-integrability, the divergence exponent $\alpha \approx 0.7$ for $\sigma = 2$ lies between the range of $\alpha = 0.3-0.5$ observed in conventional short-range nearest-neighbor systems and the value of $\alpha = 1$ characteristic of ballistic heat transport. From the perspective of nonlinear dynamics, non-integrability is generally determined by the maximum Lyapunov exponent $\lambda_{\max}$ of the system. The analysis of the maximum Lyapunov exponent in Figure 4 confirms that the non-integrability of the $\sigma = 2$ system lies between that of an integrable Toda system ($\lambda_{\max} = 0$) and a completely non-integrable near-short-range system ($\sigma = 8$).

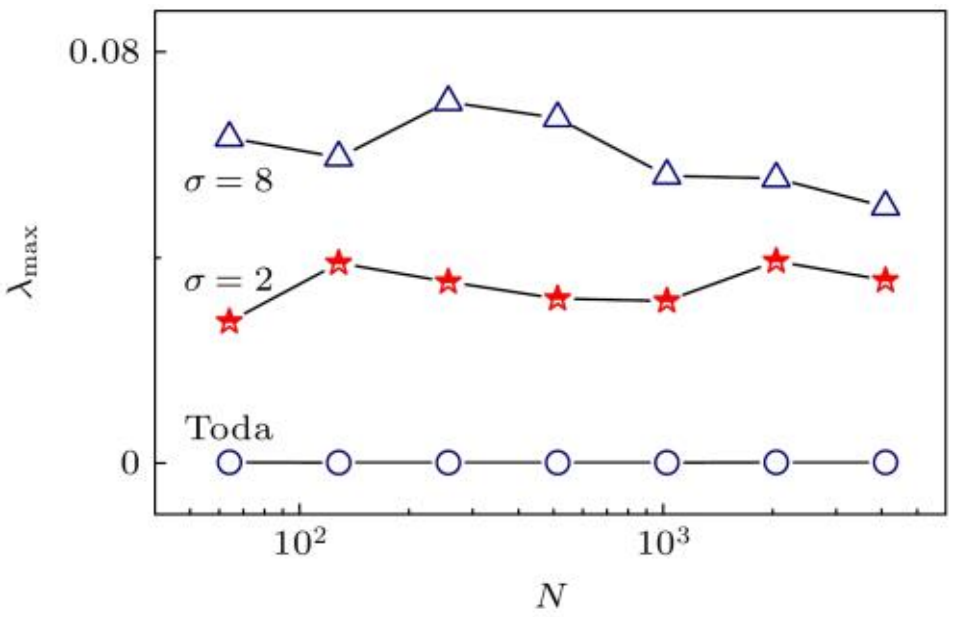

**Fig. 4** $\lambda_{\max}$ **versus system size *N* for Toda chain,** $\sigma = 2, \ 8$ **systems.**

Wang et al.[36] attribute this weak non-integrability to the unique microscopic dynamics of moving breathers within the system. Indeed, work by Doi and Yoshimura[53] in 2016 demonstrated that LR-FPUT systems with $\sigma = 2$ excite tailless moving breathers at zero temperature. This phenomenon is absent in traditional short-range systems. To determine whether these moving breathers persist at finite temperatures, the researchers analyzed energy spatiotemporal maps of the system in equilibrium steady states at finite temperatures[56]. They found that although thermal fluctuations perturb these breathers, which were theoretically proven to exist at zero temperature, they maintain robustness over extended periods. This finding further confirms the distinct nature of the $\sigma = 2$ case.

In summary, a growing consensus regarding the $\sigma = 2$ case in LR-FPUT systems holds that $\sigma = 2$ represents a significant point of enhanced transport. Its thermal conductivity exceeds the typical anomalous levels observed in short-range nearest-neighbor chains. However, whether it reaches the strict ballistic limit requires careful definition. Furthermore, non-equilibrium steady states under LR coupling exhibit heightened sensitivity to boundary conditions, heat bath coupling methods, and heat flow definitions. Consequently, phenomena such as "nearly flat temperature profiles" and "$\kappa$ increasing nearly linearly with $N$" do not constitute sufficient evidence for ballistic transport. Mechanistically, existing studies[36] tend to attribute the uniqueness of $\sigma = 2$ to the system's weak non-integrability and the enhanced energy transport facilitated by special carrier structures like moving breathers, rather than to hidden integrability. Conversely, key disagreements persist regarding whether the asymptotic thermal transport at $\sigma = 2$ approaches ballistic ($\alpha = 1$) or strong superdiffusive ($\alpha \approx 0.7$) regimes, as well as the sources of discrepancies among different numerical schemes. Therefore, cross-method consistency criteria may be necessary to define the asymptotic transport behavior closer to reality. If the divergence of $\kappa(N)$, the scaling exponent of $\rho_Q(m,t)$, and the long-time decay of $C_{JJ}(t)$ align under identical parameters such as periodic boundary conditions and remain insensitive to driving or boundary implementations, this alignment can be regarded as the asymptotic scaling of bulk properties. In contrast, if conclusions rely solely on flat temperature profiles or the apparent linear growth of $\kappa(N)$ (as in Reference [38]) without other consistent evidence, one must treat these findings cautiously and attribute them to strong boundary sensitivity.

### 3.2 Subdiffusive Thermal Transport at $\sigma = 0$

Subdiffusive motion differs from normal diffusion in that its mean squared displacement (MSD) does not increase linearly with time. Instead, it scales according to $t^{\beta}$, where $\beta < 1$. This significantly slow diffusion occurs in many complex systems. However, most studies focus only on particle motion and subdiffusion within energy transport types and their corresponding fundamental physical mechanisms is still unclear. From a more physical perspective, determining what causes energy subdiffusion and how subdiffusive energy characteristics emerge in many-body Hamiltonian systems remains a challenging topic.

Due to the finiteness of Poincaré recurrence times[57], it is generally believed that conventional Hamiltonian systems cannot achieve subdiffusive motion in the absence of disorder. Consequently, examples of subdiffusive energy transport in Hamiltonian systems are rare. The first convincing example demonstrated thermal transport with a $\beta$ value of 0.86 in a specially configured dyuamical channel model[58,59]. However, this differs significantly from many-body lattice Hamiltonian systems where phonons (i.e., collective excitation modes or quasiparticles) play the primary role. In short-range nearest-neighbor lattice systems, it was only recently discovered in 2025 that the $\phi^4$ system, which breaks momentum conservation under strong nonlinearity, can exhibit subdiffusive energy transport[60]. Notably, Bagchi[39] provided indications of subdiffusive transport in a 1D Hamiltonian mean-field LR-rotor model in 2017. However, in another 2017 study[38], Bagchi reported numerical results indicating continued superdiffusive transport in the Hamiltonian mean-field case of the LR-FPUT model ($\sigma = 0$). This controversy prompted Xiong and Wang[44] to carefully examine thermal transport in long-range FPUT systems at $\sigma = 0$. They found that, unlike Bagchi's work using NEMD methods[38], the LR-FPUT system at $\sigma = 0$ resembles the Hamiltonian mean-field LR rotor model and also exhibits energy subdiffusion. This is evidenced by the localized distribution function of the spatiotemporal correlation function of energy density fluctuations. Furthermore, its mean squared displacement shows clear characteristics of $\beta < 1$ over time (see Figure 5).

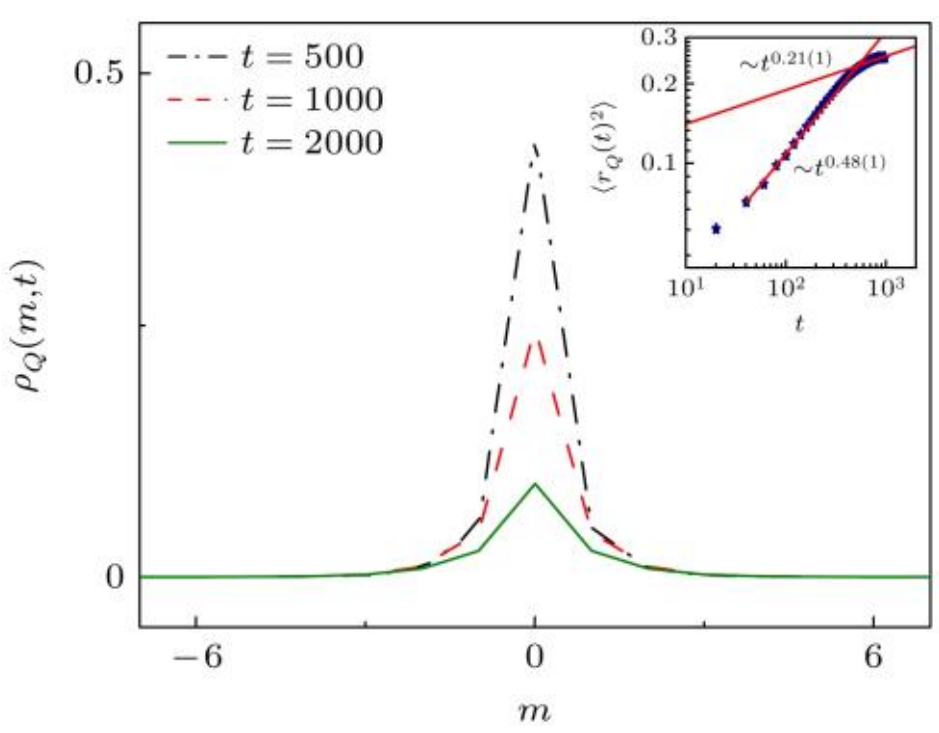

**Fig. 5 The $\rho_Q(m,t)$ of the $\sigma=0$ system at different times, the inset shows the relationship between the mean square displacement and time.**

More intriguingly, the temporal evolution of the energy current autocorrelation function exhibits distinct characteristics (see Figure 6). Xiong and Wang[44] first observed a negative correlation effect in the energy current autocorrelation function, similar to that seen in particle subdiffusion (where it corresponds to the negative correlation of the particle velocity autocorrelation function). In the academic community, this negative correlation effect is also termed antipersistent correlation. In fact, subdiffusive energy transport implies that $\kappa$ should approach zero in the thermodynamic limit, indicating that the system acts as a thermal insulator. To support this, mathematically, $C_{JJ}(t)$ must change sign at least once for the Green-Kubo integral to vanish. From a physical perspective, within the framework of particle subdiffusion models, this phenomenon signifies antipersistent characteristics, meaning that periods of positive heat flow energy transfer are typically followed by periods of negative current energy transfer. Therefore, this newly discovered antipersistent correlation in energy subtransport demonstrates that many-body nonlinear Hamiltonian systems with LR interactions can exhibit physical behaviors as rich as those in single-particle diffusion.

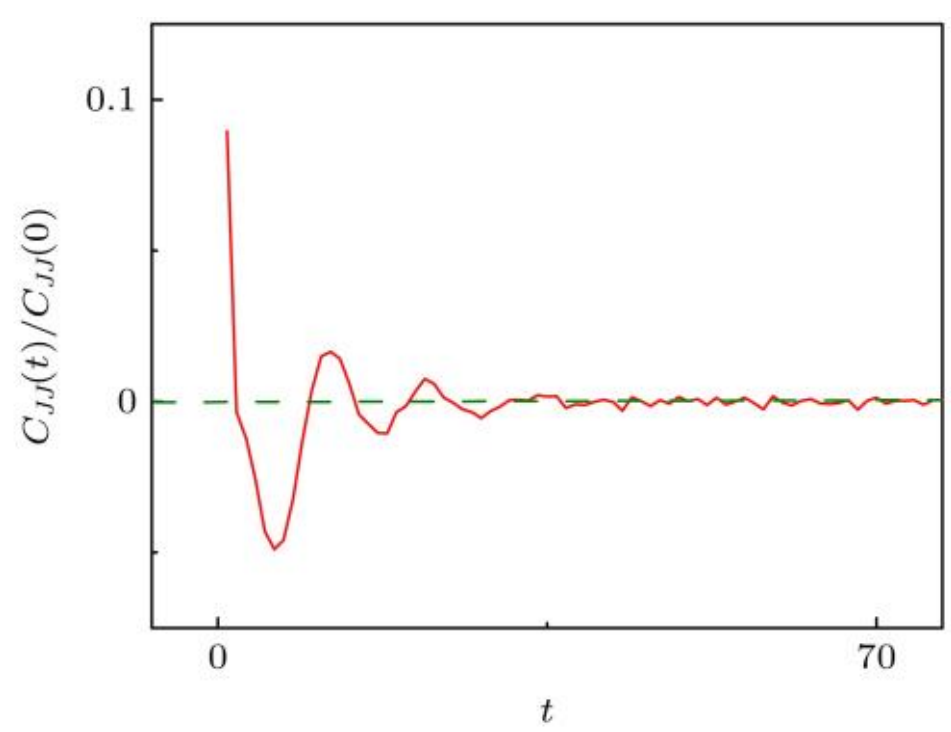

**Fig. 6 The energy flux auto-correlation function of the $\sigma=0$ system.**

However, it is crucial to emphasize that the persistent negative correlation discussed here is primarily a numerically observable correlational structure. In many subdiffusive regimes (or where effective thermal conductivity saturates or decays with scale), interpreting the thermodynamic limit via the standard Green-Kubo formalism often requires $C_{JJ}(t)$ to exhibit specific sign structures (such as a transition from positive to negative correlation) to facilitate effective cancellation or convergence of the integral. Nevertheless, in finite systems with finite time truncation and under varying heat flux definitions or boundary conditions, this relationship should not be construed as a strict necessary or sufficient condition. Therefore, this study treats negative correlation as a significant concomitant feature of subdiffusive transport. It underscores the need for cross-validation using indicators such as the localization profile of $\rho_Q(m,t)$ and the mean squared displacement (MSD) exponent $\beta < 1$.

To confirm that the subdiffusive energy transport characteristics associated with negative energy flux correlation are indeed linked to system LR interactions, Reference [44] further investigated the variation of $C_{JJ}(t)$ in models with truncated LR interaction ranges. The study found that when the LR characteristics of the system are insufficient to support subdiffusive energy transport, the negative correlation disappears. Instead, the system reverts to the long-time tail phenomenon previously observed in superdiffusive transport (see Figure 7).

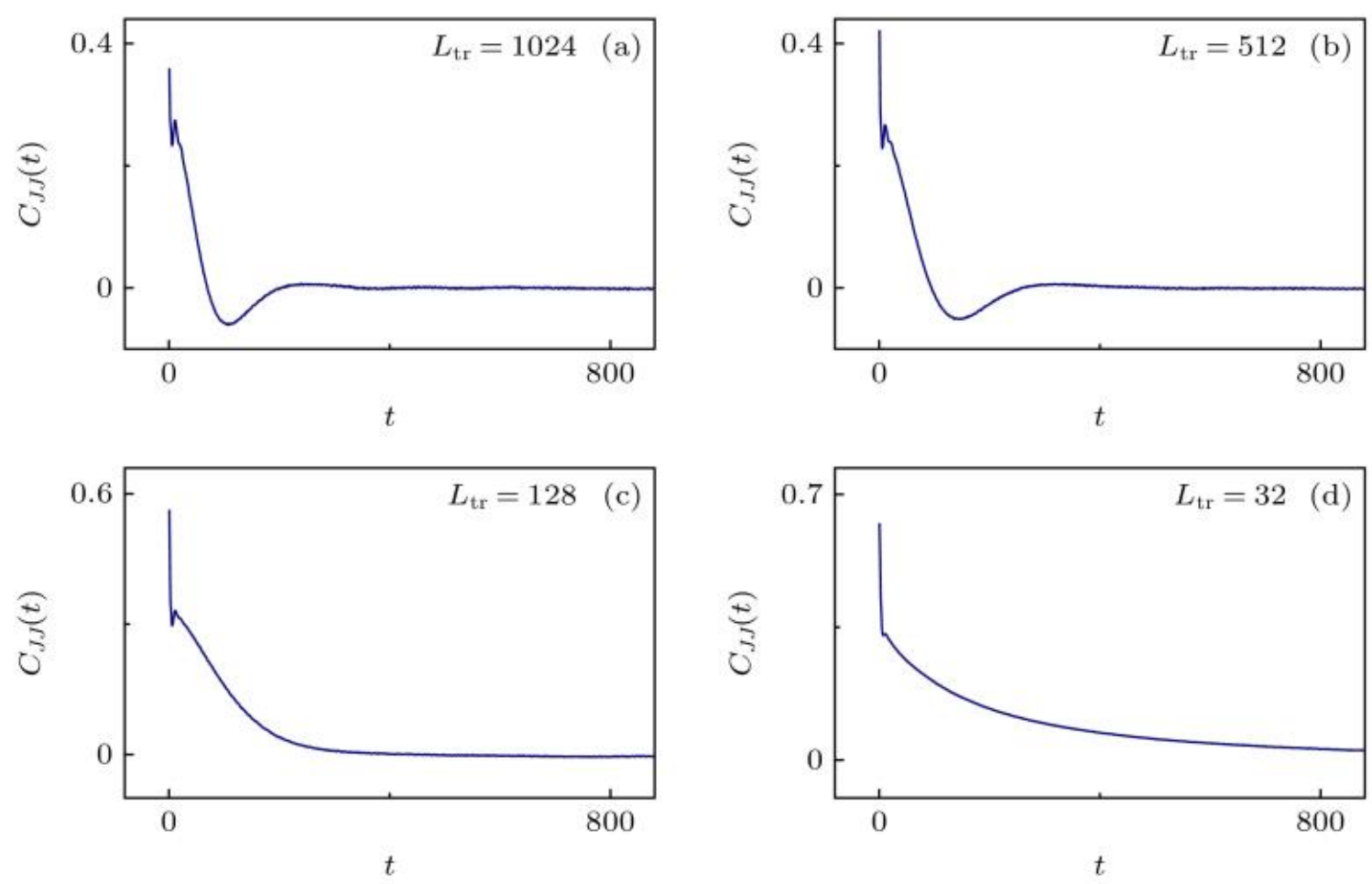

**Fig. 7 The energy flux auto-correlation function of the $\sigma = 0$ system with the long-range interaction range truncated, where $L_{\text{tr}}$ represents the actual long-range interaction length.**

### 3.3 Non-monotonic Heat Transport in the Strong Long-Range Region ($0 < \sigma < 1$)

In 2017, Bagchi[38] conducted a detailed numerical study on the variation of the divergence exponent $\alpha$ with $\sigma$ in LR FPUT systems. In addition to identifying the special case at $\sigma = 2$, the study indicated that superdiffusive heat transport, characteristic of nearest-neighbor coupling, occurs for all other values of $\sigma$. Specifically, as noted in Section 3.2, Bagchi observed superdiffusive transport at $\sigma = 0$. However, the work of Xiong and Wang[44] demonstrated that $\sigma = 0$ actually corresponds to subdiffusive energy transport. This inconsistency necessitates a detailed investigation into the variation of heat transport with $\sigma$. To this end, Xiong and Wang[44] employed the equilibrium method of calculating the autocorrelation function of energy flux to provide a detailed analysis of $C_{JJ}(t)$ in the strong LR region ($0 < \sigma < 1$). They found that $C_{JJ}(t)$ undergoes a non-monotonic change as $\sigma$ increases. For $0 \leqslant \sigma \leqslant 0.8$, while anti-persistent negative correlations persist in $C_{JJ}(t)$, the negative minimum value first decreases, reaching its lowest point at $\sigma_c = 0.5$, and then begins to increase. At approximately $\sigma = 0.8$, $C_{JJ}(t)$ exhibits a final form resembling exponential decay, indicating the absence of anti-persistent negative correlations (see Figure 8).

It is worth noting that the transition from the presence to the absence of anti-persistent negative correlations, as revealed by the non-monotonic variation of $C_{JJ}(t)$, is a novel and interesting phenomenon. This differs significantly from anti-persistent negative correlations revealed by truncating LR interactions in systems. Such variations in velocity anti-persistent negative correlations have not previously been explored in particle subdiffusion models. Therefore, this finding may provide new insights for further research on particle subdiffusion. Finally, it is important to emphasize that these anti-persistent energy flux correlation structures are a prominent numerical feature of the strong long-range region. In many known subdiffusive dynamic frameworks, negative correlations often accompany slow transport or backflow effects. Thus, they can be regarded as important accompanying indicators of subdiffusive tendencies. However, it is more prudent to avoid treating them as strict sufficient or necessary conditions. Based on this, the determination of "subdiffusion" in this interval should rely on direct transport indicators, such as the localization morphology of $\rho_Q(m,t)$ and its MSD exponent.

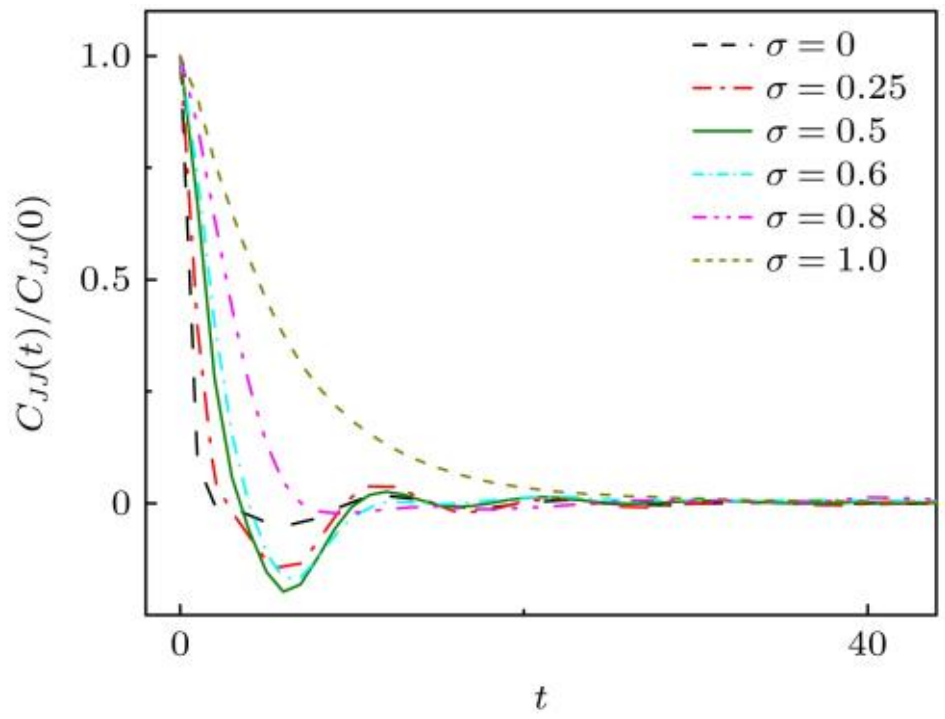

**Fig. 8 The energy flux autocorrelation functions of various systems with $0 < \sigma < 1$.**

Nevertheless, these anti-persistent energy flux correlations clearly indicate subdiffusive energy transport behavior. They strongly suggest that, similar to mean-field models, systems with strong long-range interactions should exhibit subdiffusive energy transport.

It is also worth noting a special point at $\sigma_c = 0.5$ within the strong LR regime of LR systems. This finding is somewhat unexpected, as previous understanding identified only two special points: $\sigma = 1$ (the boundary between strong and weak long-range regimes) and $\sigma = 2$ (weak non-integrable dynamics). This discovery may be closely related to chaos suppression and universal dynamic thresholds in systems with correlated LR interactions. It also provides a new direction for future research on lattice systems with LR interactions.

Furthermore, the results in Figure 8 suggest that the newly discovered $\sigma_c = 0.5$ appears to be a critical point for a dynamic phase transition. However, current literature on heat conduction lacks systematic studies on finite-size scaling and critical behavior at this boundary. Therefore, it remains unconfirmed whether these boundaries constitute strict dynamic phase transitions. Consequently, this study temporarily regards them as potential dynamic crossover behaviors or dynamic threshold phenomena.

### 3.4 Normal Heat Transport Window in the Weak Long-Range Regime ($1 \leqslant \sigma \leqslant 3$)

The academic community generally defines the weak long-range regime as $1 \leqslant \sigma \leqslant 3$. Based on this understanding, a careful examination of this regime allows further observation of how transport behavior transitions from subdiffusion to superdiffusion. Since normal diffusion lies between energy subdiffusion and superdiffusion, researchers naturally expect to observe normal heat transport. Additionally, although the specific heat conduction divergence exponents in the weak LR regime are not unified (due to differences in system size, statistical windows, heat flux definitions, and boundary/thermal bath implementations, which may cause deviations in the quantitative characterization of $\alpha$), existing numerical results[38,40] consistently indicate superdiffusive transport qualitatively. Specifically, for $1 \leqslant \sigma \leqslant 2$, Bagchi[38] reported a divergence exponent of $\alpha = 0.3$, whereas Iubini et al.[40] reported $\alpha = 0.6$. For $2 \leqslant \sigma \leqslant 3$, both Bagchi[38] and Iubini et al.[40] provided quantitatively consistent results, showing a gradual transition from $\alpha \approx 1$ to $\alpha \approx 0.4 - 0.5$. Overall, normal heat conduction satisfying Fourier's law does not appear to have been observed.

However, Xiong and Wang[45] conducted a detailed study in 2024 on the scaling laws of the heat flux autocorrelation function $C_{JJ}(t)$ and the spatiotemporal correlation function of heat energy density fluctuations, yielding unexpected results. Figure 9 shows the short-time behavior of $C_{JJ}(t)$. In the range $0.8 \leqslant \sigma \leqslant 3$, the decay is faster than $C_{JJ}(t) \sim t^{-1}$, which is a potential indicator of normal heat transport. A more precise examination of $\rho_Q(m,t)$ for each $\sigma$ reveals that normal heat transport is highly likely to occur near $\sigma = 1.25$. At this point, $\rho_Q(m,t)$ closely approximates a Gaussian distribution. Corresponding scaling analysis yields $1/\gamma = 0.53$ (see Figure 10).

Considering numerical errors, this value is very close to the theoretical value of $1/\gamma = 1/2$ for normal heat transport in Lévy walks. Furthermore, according to Lévy walk theory[55], the precise scaling exponent can be determined by the temporal decay scaling of the central peak $\rho_Q(0,t)$ of $\rho_Q(m,t)$. The corresponding detailed examination for each $\sigma$, as shown in Figure 11, confirms this finding.

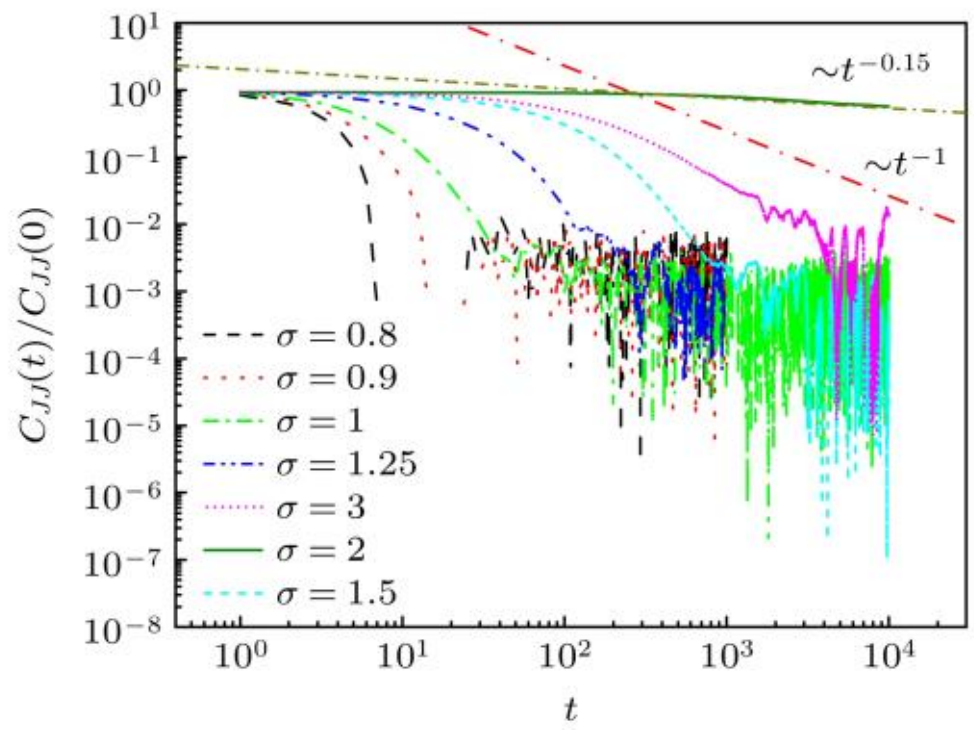

**Fig. 9 The energy flow autocorrelation functions of various systems in the weak long-range region.**

Regarding the window of normal heat transport identified in the weak LR regime ($1 \leqslant \sigma \leqslant 3$), it is worth noting that the initial theoretical motivation for studying heat transport was to investigate the microscopic factors and origins of normal heat transport satisfying Fourier's law of conduction. Previous decades of work focused on lattice systems with short-range nearest-neighbor interactions. Within that framework, researchers never observed normal thermal conduction in momentum-conserving lattices. Consequently, the discovery of normal transport centered around $\sigma = 1.25$ suggests that insights derived from short-range nearest-neighbor lattice systems require comprehensive revision. This finding also demonstrates that the physics of heat transport in systems with LR interactions is richer.

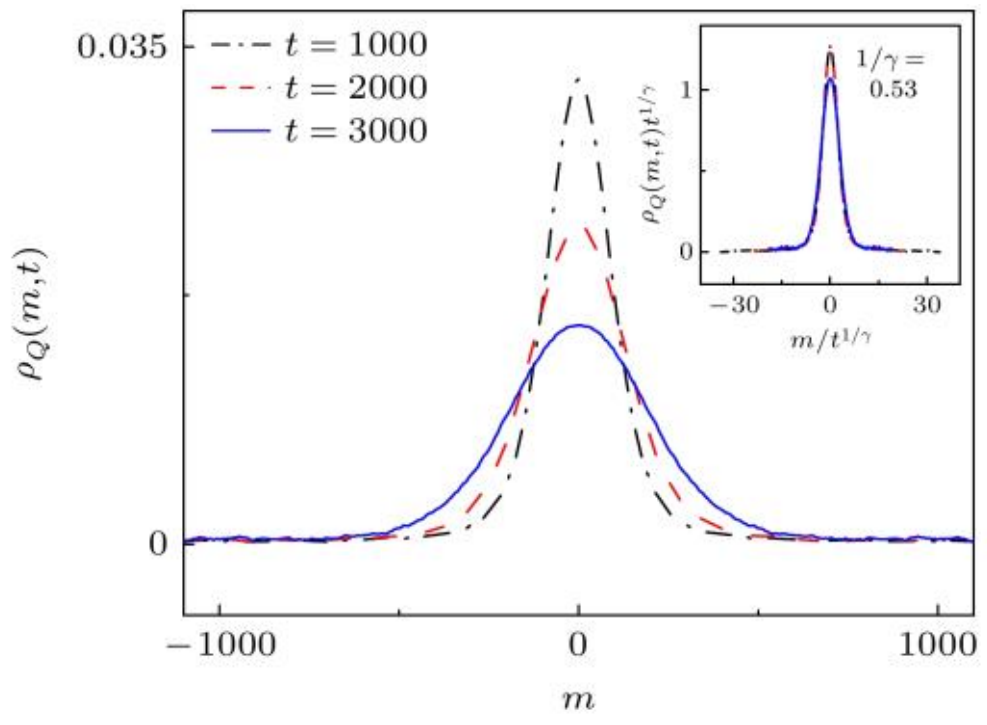

**Fig. 10 The $\rho_Q(m,t)$ of the $\sigma = 1.25$ system at different times, the inset shows its scaling behavior.**

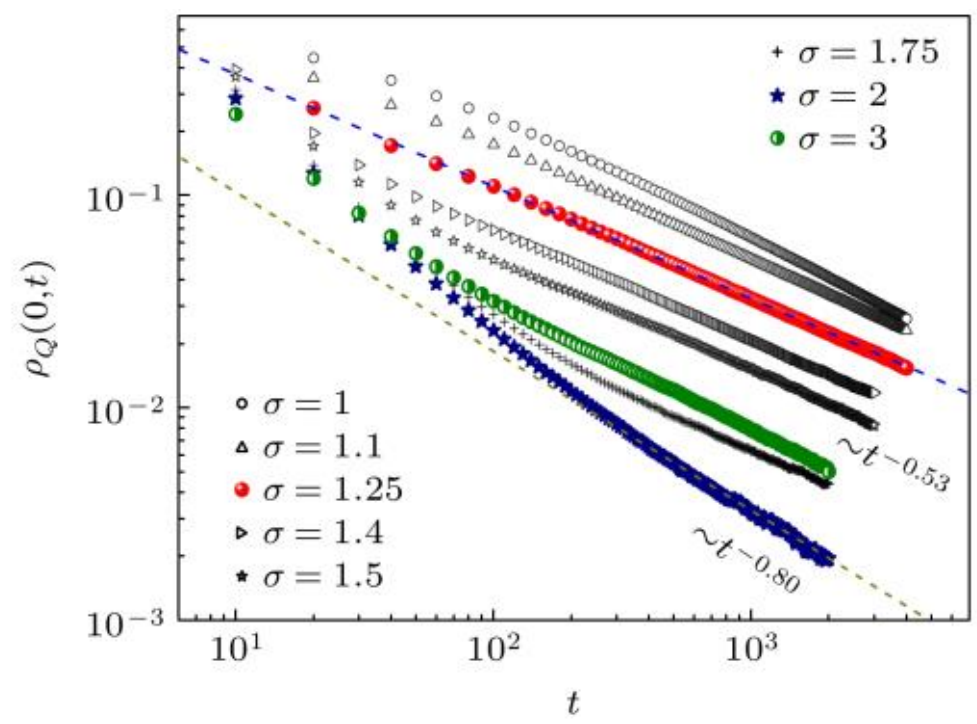

**Fig. 11 The time decay of the height of the central peak of $\rho_Q(m,t)$ of various systems in the weak long-range region.**

In traditional 1D nonlinear lattices with short-range interactions that conserve momentum, existing theoretical and numerical studies have established a robust "empirical picture": thermal conductivity diverges as a power law with system size, and Fourier's law fails in the strict 1D limit. Consequently, momentum conservation is often regarded as a key prerequisite for anomalous heat conduction. In contrast, the emergence of a near-diffusive window around $\sigma \approx 1.25$ in the weak LR regime carries significant physical implications. It suggests that introducing power-law long-range anharmonic couplings, while preserving Hamiltonian dynamics and momentum conservation, may open a mechanistic pathway toward (quasi-)normal transport. This finding challenges the conventional view that momentum conservation inevitably leads to anomalous heat conduction. Therefore, this paper summarizes the conclusion as follows: a quasi-normal heat conduction window with near-diffusive behavior emerges in the weak LR regime, implying that LR interactions may provide a new mechanistic channel for realizing Fourier's law in momentum-conserving systems.

It is important to emphasize that the current description of this "near-diffusive window" is more appropriately characterized as a numerical observation of quasi-normal or near-diffusive behavior. Its challenge to existing paradigms stems from the consistent alignment of multiple indicators, while its "pending confirmation" arises from the numerical difficulties inherent in LR systems.

On one hand, from the perspective of the equilibrium energy current autocorrelation function, $C_{JJ}(t)$ exhibits significant differences for different $\sigma$ values in the weak long-range regime. Specifically, the curve for $\sigma = 1.25$ shows decay characteristics in the long-time tail that are closer to the diffusive case. On the other hand, a review of the literature reveals significant discrepancies in the thermal conductivity divergence exponent $\alpha$ across different $\sigma$ intervals in the weak long-range regime (for example, quantitative differences in $\alpha$ within the range $1 \leqslant \sigma \leqslant 2$). Beyond finite-size effects and statistical windows, these discrepancies may also be influenced by differences in the definition of heat flux and the implementation of boundaries or thermal baths in long-range systems. Therefore, judgments regarding "strictly normal" behavior must be made with greater caution.

Consequently, this paper adopts a more conservative stance: the results for $\sigma = 1.25$ are regarded as a candidate window worthy of close attention. We emphasize the need for more systematic cross-validation under unified numerical settings. For instance, one should simultaneously compare the non-equilibrium $\kappa$ scaling, the long-time tails of the Green-Kubo formula, and the spatiotemporal scaling of energy density fluctuations to rule out the possibility of "finite-size crossover artifacts." Furthermore, this paper points out that the current literature lacks systematic scans and finite-size scaling analyses for several characteristic $\sigma$ points (including $\sigma = 1.25$) at large system sizes. Thus, it remains unconfirmed whether these boundaries represent continuous crossovers or dynamical phase transitions.

From a theoretical perspective, the conclusion that "momentum conservation leads to anomalous heat conduction" in short-range systems is often based on hydrodynamic descriptions induced by local conservation laws and short-range interactions. However, in the presence of LR non-local couplings, the long-time memory kernels of transport, effective scattering channels, and mode-coupling structures may undergo qualitative changes, necessitating a re-examination of traditional short-range empirical rules. Therefore, the "near-diffusive window" at $\sigma \approx 1.25$ does not imply an overturning of existing understanding. Rather, it suggests that LR non-locality provides ID systems with an "effective relaxation pathway/scattering landscape" distinct from short-range scenarios. Its asymptotic limits and universality require further clarification.

### 3.5 Ballistic Heat Transport in Non-integrable Systems with Staggered Long-Range Coupling

In 2022, Yoshimura et al.[61] modified the traditional LR-FPUT lattice system into a staggered long-range coupled FPUT-like lattice system. They claimed that this system exhibits the disappearance of phonon umklapp processes. The Hamiltonian of this system is

$$H = \sum_{i}^{N}\{\frac{p_i^2}{2} + V_{\mathrm{NN}}(x_{i+1} - x_i) + \sum_{r=1}^{N/2-1} V_{\mathrm{LR}}[x_{i+r} - (-1)^r x_i]\}, \quad (6)$$

In this equation, the interparticle interaction consists of two terms: the nearest-neighbor harmonic term $V_{\mathrm{NN}}(\xi) = \xi^2/2$ and the long-range anharmonic term $V_{\mathrm{LR}}(\eta) = \frac{k_{\mathrm{LR}}\eta^4}{4r^2}$ (where $\eta_i = x_{i+r} - (-1)^r x_i$, and $k_{\mathrm{LR}}$ characterizes the nonlinear strength of the system); $(-1)^r$ reflects the staggered coupling of the long-range term.

According to Peierls theory, the absence of phonon Umklapp processes implies zero thermal resistance. Consequently, the system exhibits ballistic heat transport, a phenomenon typically observed only in integrable systems within the framework of nearest-neighbor short-range interactions. Due to computational challenges, Yoshimura et al.[61] employed traditional NEMD methods to approximate a LR system by increasing a cutoff distance for long-range interactions under fixed boundary conditions. At low nonlinear strength ($k_{\mathrm{LR}}$ = 0.1), their data revealed a flat temperature profile and thermal conductivity that increased linearly with system size. From the perspective of theoretical developments in heat transport in low-dimensional systems, this work broke through the conventional understanding established for short-range systems, which held that only integrable systems could exhibit ballistic heat transport. Their study demonstrated that introducing appropriate long-range interactions allows a non-integrable system to exhibit ballistic transport, at least mathematically. However, Yoshimura et al.[61] restricted their analysis to low nonlinear strength, which easily falls within the regime of linear integrable systems in finite systems. Furthermore, as previously mentioned, studying heat transport in long-range systems using traditional NEMD methods inevitably encounters difficulties related to boundary and heat source coupling. Therefore, their work was limited to discussing the construction and verification of a system with vanishing phonon Umklapp processes.

In 2025, Wang and Xiong[50] extended this system with vanishing phonon Umklapp processes from the near-linear regime ($k_{\mathrm{LR}}$ = 0.1) to the fully nonlinear regime ($k_{\mathrm{LR}}$ = 1). They systematically calculated all characteristic signatures of ballistic transport, such as a flat bulk temperature profile (see Figure 12), the ballistic scaling law of spatiotemporal correlations in thermal energy density fluctuations, the ballistic growth of mean square displacement over time as indicated by the transport distribution function (see Figure 13), and the non-decaying energy current autocorrelation function at long times (see Figure 14). This study provided robust data supporting ballistic heat transport in non-integrable systems for the first time. It demonstrated that ballistic heat transport is feasible in fully chaotic non-integrable systems, indicating that "ballistic transport is not necessarily equivalent to integrability." Thus, this work comprehensively enriches the overall understanding of heat transport in both LR and short-range systems.

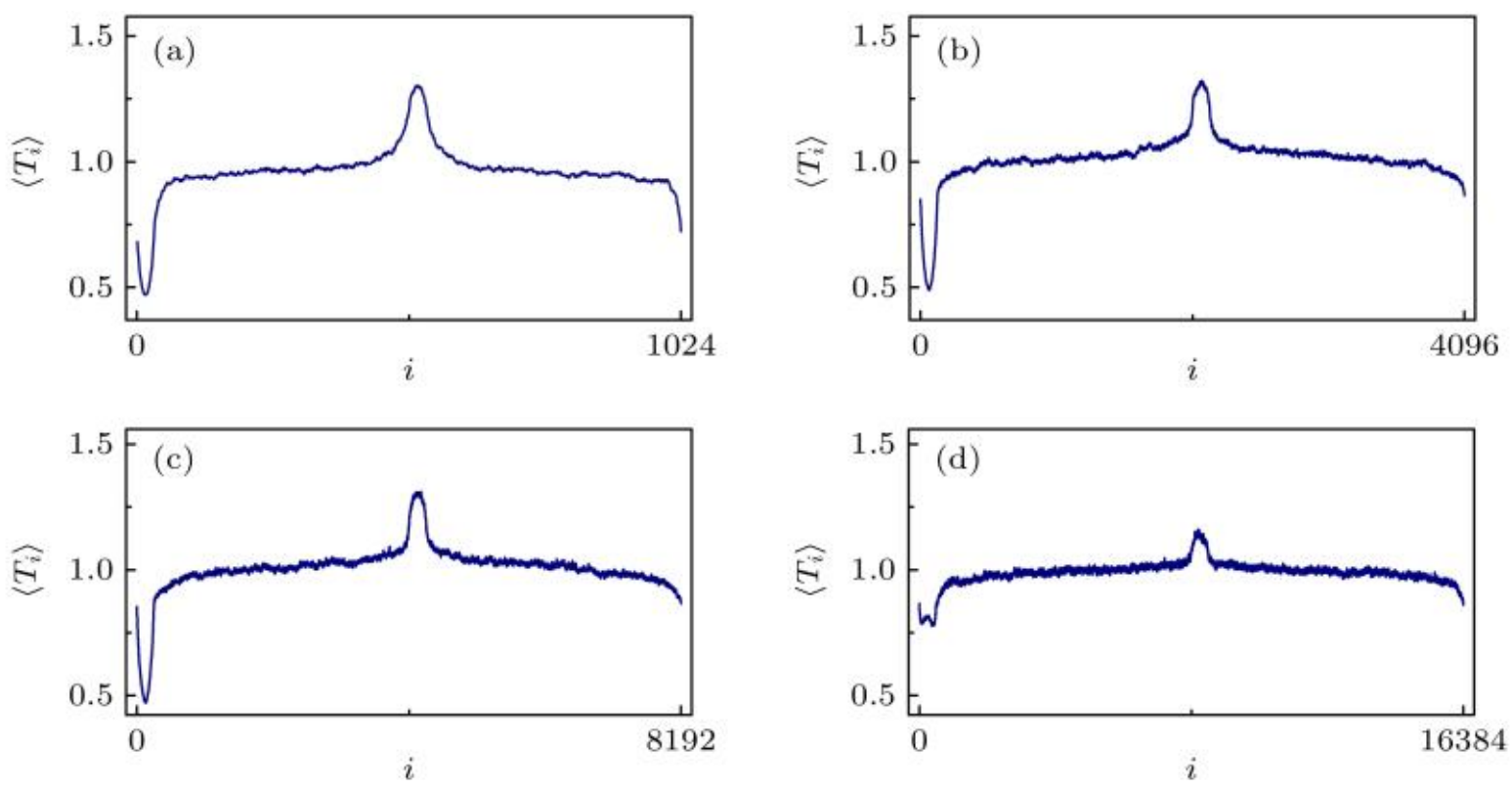

**Fig. 12 Temperature distribution of the staggered long-range coupled non-integrable system at different system sizes.**

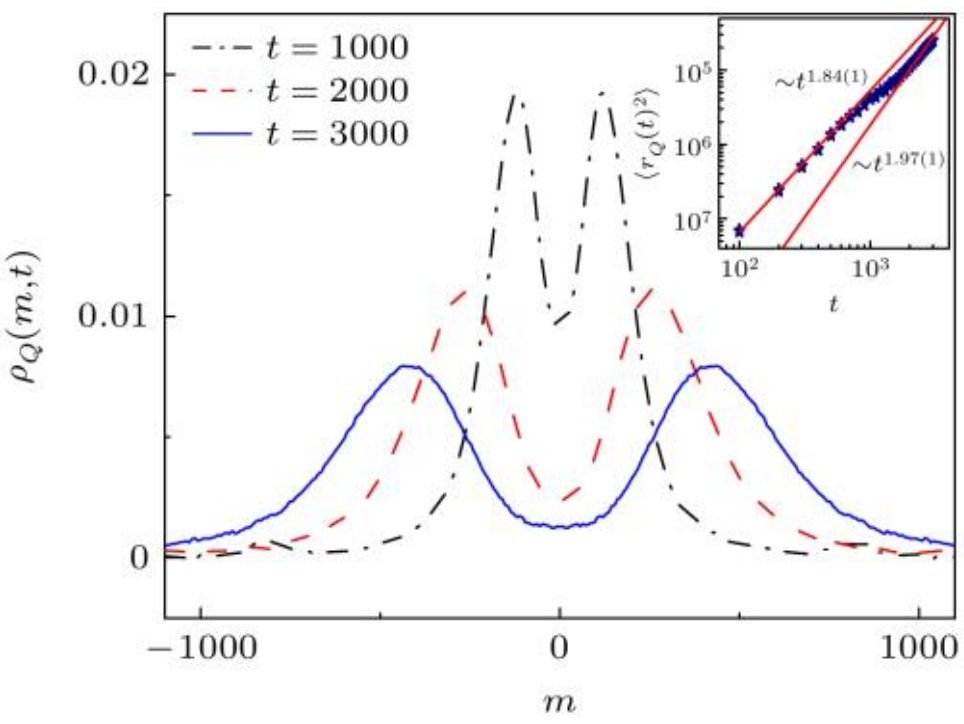

**Fig. 13 The $\rho_Q(m,t)$ of the staggered long-range coupled non-integrable system at different times, the inset shows the relationship between its mean square displacement and time.**

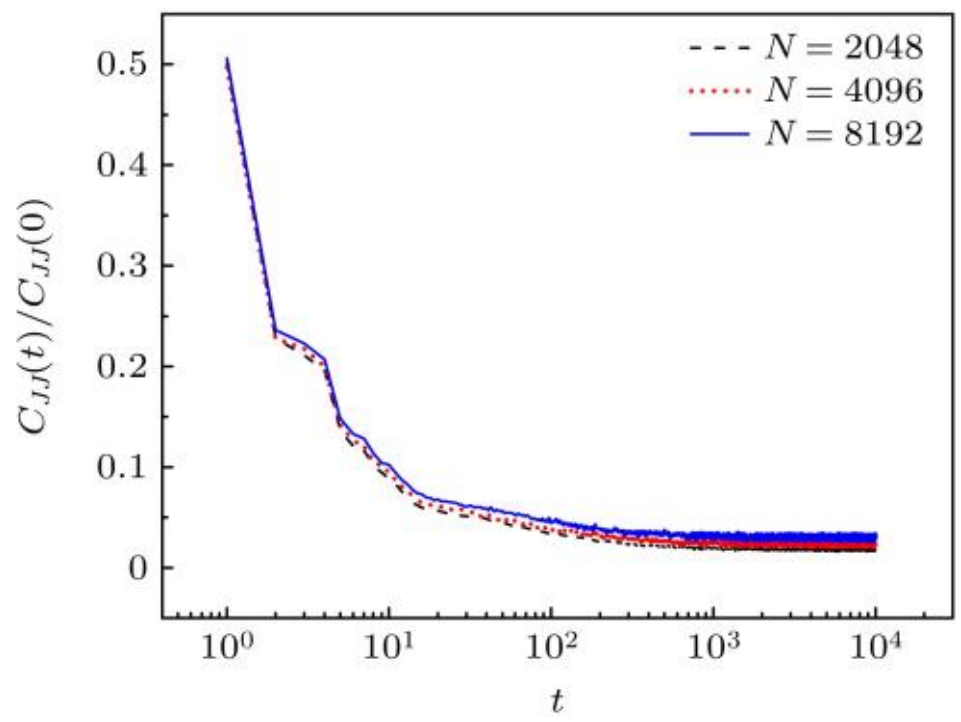

**Fig. 14 The energy current cuto-correlation function of the staggered long-range coupled non-integrable system.**

# 4 Conclusion and Outlook

This paper systematically reviews recent advances in nonequilibrium heat transport within one-dimensional LR interacting FPUT-like lattices. Under the constraints of Hamiltonian dynamics and momentum conservation, extending interactions from nearest-neighbor to power-law decaying LR forms is sufficient to induce novel transport phenomena distinct from short-range systems. These phenomena include a significant enhancement of the thermal conductivity divergence exponent, antipersistent negative correlations in heat flux autocorrelation, a normal heat transport window in the weak long-range regime, and ballistic transport in non-integrable systems. Specifically, this paper summarizes the following key findings.

1) In LR-FPUT systems with a specific interaction decay exponent of $\sigma = 2$, the divergence exponent of thermal conductivity with system size is significantly enhanced. This results in a scaling regime characterized as "quasi-ballistic but still superdiffusive." Evidence from the maximum Lyapunov exponent and energy spatiotemporal maps attributes this behavior to weak non-integrability and specific moving breather dynamics.

2) In the mean-field limit and strong long-range regime ($0 \leqslant \sigma \leqslant 1$) of LR-FPUT systems, energy transport transitions from superdiffusion to subdiffusion. This transition is accompanied by antipersistent negative correlation structures in the heat flux autocorrelation function and their non-monotonic evolution with $\sigma$. This indicates that LR interactions not only alter power-law exponents but may also fundamentally rewrite the carrier scattering landscape.

3) In the weak LR regime ($1 \leqslant \sigma \leqslant 3$) of LR-FPUT systems, a normal heat conduction window approaching normal diffusion is observed near $\sigma \approx 1.25$. The thermal energy density fluctuation correlation function approaches a Gaussian distribution, and the scaling exponent approximates the normal diffusion value. This suggests that LR interactions may open new mechanistic pathways for realizing Fourier's law in momentum-conserving systems.

4) Complete characteristics of ballistic transport can be observed in appropriately constructed staggered long-range coupled non-integrable systems. This demonstrates that ballistic transport is not necessarily equivalent to integrability. LR coupling may achieve resistance-free transport in non-integrable backgrounds by suppressing certain key scattering or umklapp processes.

Based on these advances, this paper proposes the following open questions for further consideration.

1) To our knowledge, current literature lacks systematic studies on the critical behavior of the aforementioned $\sigma$-driven transport partitions. In particular, finite-size scaling analyses across various system sizes $N$ are absent. Consequently, it remains unconfirmed whether these boundaries constitute strict dynamical phase transitions. Currently, evidence for characteristic points such as $\sigma = 0.5$, $\sigma \approx 1.25$, and $\sigma = 2$ relies primarily on finite-scale numerical observations. Systematic finite-size scaling and critical exponent studies are still lacking. Future work requires systematic scanning of the effective exponent $\alpha_{\text{eff}}(\sigma; N)$, heat flux negative correlations, and the scaling behavior of $\rho_Q(m, t)$ under unified boundary conditions and sufficiently large $N$. This will help determine whether these boundaries represent continuous crossovers or dynamical phase transitions and explore potential universality classes.
2) This paper primarily focuses on results under fixed average temperature (e.g., $T = 0.5$) and fixed nonlinearity strength. In short-range FPUT chains, the heat transport scaling exponent $\alpha$ is known to vary with temperature. Investigating how temperature and nonlinearity strength influence the various transport phenomena described herein, and whether a phase diagram exists, represents an important direction for future research.
3) A core conclusion of this paper is that extending nonlinear interactions from short-range to power-law LR forms in 1D momentum-conserving lattices can induce diverse behaviors. These include strong superdiffusion enhancement, a near-normal diffusion window in the weak LR regime, antipersistent heat flux correlations and even subdiffusion in the strong LR regime, and ballistic transport under specific structural designs. These phenomena suggest that LR interactions are not merely parameter perturbations altering exponents. Instead, they modify the spatial structure of energy exchange and long-time memory at the microscopic level. This directly challenges existing theoretical frameworks, particularly nonlinear fluctuating hydrodynamics theory developed for nearest-neighbor short-range systems[15,16]. Testing the applicability of fluctuating hydrodynamics theory in LR-FPUT systems and developing heat conduction theory for the hydrodynamic regime in LR systems are important future research directions.
4) This review primarily presents novel transport phenomena in 1D LR-FPUT systems in recent years. What are the microscopic dynamical mechanisms underlying these phenomena? Doi and Yoshimura[53] proved the existence of tail-free, non-decaying moving breathers in LR-FPUT systems with $\sigma = 2$ at zero temperature. Wang et al.[36] first linked superdiffusive transport at $\sigma = 2$ to moving breathers discovered in the system. Xiong and Wang[45] identified special stationary breather configurations at $\sigma = 0.5$. Xiong and Wang[44,50] associated complex transport variations with breather and phonon scattering. However, these attempts represent only preliminary phenomenological associations. A more in-depth study of different carriers and their scattering dynamics across various systems remains a direction worthy of detailed exploration.
5) In recent years, deeper insights have been gained into classical thermalization in short-range nearest-neighbor systems. Notably, Professor Zhao Hong's research group at Xiamen University discovered a universal scaling law where thermalization time is inversely proportional to the square of perturbation strength[20-24]. However, research on thermalization in LR interacting systems remains in its preliminary stages (e.g., work by Professor Wang Jian's group at Yangzhou University[62]). Whether this universal scaling law discovered in short-range systems also applies to LR interacting lattice systems is a question worthy of investigation.

6) Finally, it is worth emphasizing that LR interactions are not purely theoretical constructs. Existing experimental techniques can prepare and manipulate real materials and artificial structures with LR coupling. For instance, systems such as Coulomb crystals[63], Ising pyrochlore magnets[64,65], and permalloy nanomagnets[66] naturally or effectively exhibit LR interaction characteristics. More importantly, high-efficiency thermal rectification has been achieved in systems with LR interactions in recent years[35,67,68]. This demonstrates the practical feasibility of controlling heat flow directionality and magnitude through interaction range engineering. Combining the review results of ID LR FPUT-like lattices presented herein, we suggest the following implications for understanding and controlling thermal properties of actual materials: Long-range coupling not only alters the scaling laws of traditional phonon scattering and energy diffusion but may also introduce richer dynamical structures under nonequilibrium conditions. Especially under the combined action of "staggered long-range coupling + nonlinearity," the system can present a complex dynamical landscape where chaotic and regular motions intertwine. This mechanism of controlling heat flow via phase space structure provides a new perspective for explaining and designing thermal rectification effects. In other words, beyond enhancing scattering or introducing disorder, the coherent/incoherent competition induced by the synergy of LR coupling and nonlinearity, along with the resulting reconstruction of heat flux correlations and transport channels, may become key physical handles for constructing high-performance thermal diodes, thermal switches, and thermal logic devices. This could promote the leap from model research to the design of controllable heat transport materials and devices.